
\magnification=\magstep1
\line{\hfill UMD 93-038}
\line{\hfill gr-qc/9209009}
\bigskip
\centerline{ EUCLIDEAN MAXWELL-EINSTEIN THEORY\footnote*{to appear
 in Louis Witten Festschrift (World Scientific)}}
\bigskip
\centerline{ DIETER BRILL}
\centerline{\it Department of Physics}

\centerline {\it  University of Maryland}

\centerline{\it College Park, MD 20742, USA}
\bigskip \bigskip

\noindent {\bf Abstract}

After reviewing the context in which Euclidean propagation is useful
we compare and contrast Euclidean and Lorentzian Maxwell-Einstein theory
and give some examples of Euclidean solutions.

\bigskip
\noindent {\bf 1. Introduction}
\medskip
The sourcefree Maxwell equations minimally coupled to Einstein gravity
represent a classical field theory to which L. Witten has made important
contributions$^1$. Today this theory is still of considerable interest,
not least because it contains features not present in pure Einstein
gravity; for example, certain aspects of black holes can be simpler in
Maxwell-Einstein than in pure Einstein theory. As we are today groping
toward quantization of gravity, the classical equations and their solutions
can give us new insights when considered in their imaginary-time or
Euclidean form, because such solutions can be used as the basis for a WKB
approximation to the as yet unknown quantum theory.

In the realm of gravitational physics, Euclidean solutions have been
considered primarily in pure Einstein gravity and its higher-dimensional
versions on the one hand, and in string theory on the other.
But because we have
a much better intuitive and practical understanding of electromagnetism
than of other fields, it is appropriate to consider Maxwell-Einstein theory
as a model, and it is interesting in particular to compare and contrast
Euclidean electrodynamics with the standard (Lorentzian) version.
That the differences are small and subtle is suggested by the fact that
Maxwell's equations are metric-independent if written in terms of two
2-forms (one containing $B$, $E$, the other $D$, $H$). The metric enters only
in the ``vacuum constitutive relations," which demand that these tensors
are duals of each other.

We first recall the relation between Euclidean time development and the
WKB approximation in classically forbidden regions. Then we consider Euclidean
vs.~Lorentzian Maxwell theory and show that the only essential difference
is the sign in Lenz' law. We give examples of solutions that illustrate
this difference, and comment on the peculiar role of duality transformations.
We conclude by giving Einstein-Maxwell solutions that are amusing even if
their physical significance may be a matter of speculation.

\break

\noindent {\bf 2. Tunneling, Bounces and Instantons}
\medskip
The motivation for considering Euclidean theories can be explained by the
simple example of one-dimensional particle motion $x(t)$ in a potential.
In regions where the potential energy $V$ exceeds the kinetic energy $T$,
classical motion is forbidden, but quantum tunneling may take place. In
a semiclassical description of tunneling one imagines that, for a fixed
total energy, real time ``freezes" and barrier penetration occurs by the
particle ``taking an excursion into complex time."$^2$ The complex-time
motion can then be used as a basis for computing the WKB wavefunction.

The usual quadrature for the total time taken by a particle of total energy
$E = 0$ to move between $x_1$ and $x_2$,
$$t_{12} = \int_{x_1}^{x_2} \sqrt{-m/2V}\, dx  \eqno{(1)}$$
shows that $t$ will be complex if the interval between $x_1$ and $x_2$
includes a forbidden region (where $V > 0$). If we allow $t$ to vary
from $0$ to $t_{12}$ along some path (contour) in the complex $t$-plane,
there will typically be a complex solution $x(t)$ to the equations of
motion satisfying the boundary conditions $x(0) = x_1$, $x(t_{12}) = x_2$
and $E = 0$. This contour can then be deformed to consist of segments with
either purely real or purely imaginary tangent, in such a way that $x$ is
purely real.$^{2,3}$ At a (simple) zero of $V$ the contour turns abruptly by
$90^{\circ}$, so we still call these turning points (though it is the
direction of $t$ that turns, rather than the velocity of the classical
motion).

A family of such solutions corresponds in the usual way to a solution $S$ of
the (time-independent) Hamilton-Jacobi equation, and the latter gives the
``phase of the wavefunction" in the lowest-order WKB approximation. Because
the momentum $p = dS/dx$ will be imaginary in the classically forbidden
region, $S$ will generally be complex. Classically we can give no meaning to
complex $S$, but in the factor $e^{iS/\hbar}$ of the wavefunction, the
imaginary part of $S$ simply describes the exponential decrease associated
with tunneling. As usual one does not have to solve the Hamilton-Jacobi
equation to find the value of $S$, but can instead evaluate the action integral
$$I = \int \left({1 \over 2}m\dot{x}^2 - V(x) \right)dt \eqno{(2)}$$
``on shell", i.e., for the solution $x(t)$ and the same contour as described
above.

Often one is interested only in the exponential factor that the wavefunction
acquires between two adjacent turning points. One then needs to find the
imaginary-time motion between these turning points, and evaluate the action
for this motion. Although this is a needlessly arduous way to arrive at the
simple formula, $S = \int \sqrt{-2mV}\, dx$ in the one-dimensional case, it is
the procedure that can be generalized to higher dimensions. The corresponding
solutions of the equations of motion in imaginary time and with finite action
are called instantons. One can show$^3$ that at a turning point (which is a
whole spacelike surface in field theory), {\it all} momenta should vanish
(for example, because the momenta of the imaginary-time motion are imaginary,
but should agree at the turning point with those of the real motion).
Of particular importance are motions that start at a
classically static solution at $t = -\infty$, and that have precisely one
turning point. Such a motion is called a bounce\footnote*{In more precise
language than used here, bounces and instantons are mutually exclusive
categories.} and is useful, for example, in describing vacuum decay.$^4$

In an instanton we can avoid the one remaining source of imaginary quantities,
the imaginary time $t$, by making a coordinate transformation to a real
$\tau = t/i$. This changes the signature of the metric to positive definite,
making the spacetime Euclidean (or Riemannian, if there is curvature).
Covariant field equations have their old form in the new metric. Only those
quantities change their form that are by convention not ``analytically
related", i.e., not related by the above coordinate transformation.
A typical example is the volume element, $\sqrt{-g}\,d^4x$ in Lorentzian
spacetimes. If analytically continued to Euclidean it would become imaginary,
but convention defines it to be real, $\sqrt{g}\,d^4x$ in Riemannian spaces.
Another example is the length of timelike vectors. Quantities involving
such objects, such as the duality ($\ast$) operator or the extrinsic curvature
(and other quantities arising from a 3+1 split) similarly change their form,
and are not analytically related. Where this is important one invents more
or less ad hoc rules. For example, one defines the Euclidean action $S_E$
of a real Euclidean field history to be real (one uses an expression identical
to the Lorentzian action $I$, except with an overall minus sign and the
Euclidean metric and volume element) and remembers that it contributes a factor
$e^{-S_E/\hbar}$ to the wavefunction.
\bigskip
\noindent {\bf 3. The Euclidean Maxwell Equations}
\medskip
The difference between the Lorentzian and Euclidean Maxwell equations
are not apparent in their (identical) covariant form, but do emerge in
the ususal 3+1 dimensional description. Contemplation of the Euclidean
``dynamics" in this form is not only amusing, but may give at least
some measure of satisfaction to those who would like to imagine,
to the extent possible, what goes on during tunneling; at best the
good physical intuition we have about Lorentzian electromagnetism may translate
into a better understanding of Maxwell and Einstein-Maxwell instantons.

For simplicity we consider the Euclidean Maxwell equations in flat space
(but much of what we will say will apply, {\it mutatis mutandis}, to curved
space as well),
$$ {\rm div} E = 0, \quad {\rm div} B = 0, \quad
{\rm curl} E = \partial B/\partial t,
\quad {\rm curl} B = \partial E/\partial t.  \eqno{(3)}$$
The notable difference from the usual vacuum Maxwell equations is that
the minus sign, which goes by the name of Lenz' law, is absent. Thus Euclidean
electrodynamics is ``just like" ordinary electrodynamics except for an
``anti-Lenz" law. This one sign change has, however, far-reaching effects.
It changes the equations from hyperbolic to elliptic, so there is no
propagation with a finite speed in Euclidean spaces. On the other hand,
there can be run-away solutions, because now induced (displacement) currents
reinforce those that led to the induction. Thus one can have creation
of Euclidean electromagnetic fields ``from nothing".  Energy
conservation does not prevent such solutions, because Euclidean energy,
$$E = \int T_{00} d^3V = (1/8\pi) \int(B^2 - E^2)\,d^3V  \eqno{(4)}$$
is not positive definite.

Nonetheless such solutions cannot describe a decay of the vacuum: for this
we would need a proper bounce solution with a turning surface on which all
momenta, that is all $E$-fields, vanish. For such pure magnetic configurations
Eq.~(4) is positive definite. In other words, because tunneling conserves
energy, and Lorentzian electromagnetic energy is positive definite, there
is no Lorentzian state to which the vacuum could decay.\footnote*{This is of
course well known, since we do know how to quantize electromagnetism not
only in the WKB approximation, but in general. The point here is to learn
something about instantons; also we will see that the story changes in
a background field and in interaction with gravity.}

The Euclidean action of a Euclidean solution is
$$S_E = (1/16\pi) \int F_{\mu \nu} F^{\mu \nu} \sqrt{g} \, d^4x =
 (1/8\pi) \int(E^2 + B^2)\,d^4V   \eqno{(5)}$$
and is manifestly positive, as is appropriate for a barrier penetration factor
(and similarly for contributions to the path integral). It is interesting
that this action is invariant both under the Lorentzian duality
($E \rightarrow B$, $B \rightarrow -E$), and under Euclidean duality
($E \rightarrow B$,  $B \rightarrow E$), which preserves Euclidean
solutions.\footnote{$^{\dag}$}{When written covariantly, either duality
becomes$^1$ $ F \rightarrow \ast F$ but, as mentioned above, the Euclidean
$\ast$ as conventionally defined is not the analytic continuation of the
Lorentzian $\ast$. For example, $\ast \ast = +1$ for Euclidean, and $-1$
for Lorentzian spaces.}
(In both cases the invariance is valid only on shell, i.e.,
for solutions of the field equations. This is so because $E$ and $B$
cannot be freely varied, but must obey flux conservation,
$F_{[\alpha \beta ,\gamma ]} =0$, or be derived
from a potential). This assures that physical quantities, such as
barrier penetration factors, are also invariant under Lorentzian duality
of the initial state.

This last statement may seem surprising in view of our earlier statement
that $E$ must vanish on a turning surface. If an initial (or final)
Lorentzian $E$ were present, it might also seem that it should somehow
contribute with the opposite sign to $S_E$ than $E$ fields induced in
the Euclidean development.  Nonetheless Eq.~(6) is the correct expression
in either case, and for any case that can be transformed to a pure $B$
field by duality transformation (and then forms a proper instanton).
The reason is that the electromagnetic action is to be varied subject to
flux conservation; the situation is analogous to barrier penetration
in a spherically symmetric potential when there is a conserved
angular momentum.$^5$

\bigskip
\noindent {\bf 4. Examples of Maxwell-Einstein Instantons}
\medskip
In gravitational physics Euclidean solutions of interest appear to be either
tunneling ``from nothing" (universe creation), bounces (decay of a classically
static state), or instantons proper (fluctuations). Bounces have the clearest
physical interpretation, but there are not many suitable initial states:
the initial state must not only be static, but also not uniquely determined by
its energy (so there is another, non-static state to which to tunnel).
Therefore
there are no examples in pure, asymptotically flat Einstein theory.

In some theories, such as higher-dimensional compactified pure Einstein or
low-energy string theories, the positive-energy theorem does not hold. The
principal example of a bounce that signals vacuum decay in such a theory
was given by the son of L. Witten.$^6$ The analogous decay with an
electromagnetic field exists in various dimensions.$^7$ To illustrate Euclidean
Maxwell theory we consider the four-dimensional case, which is somewhat
unrealistic because one dimension must be compactified. So the initial state
is the gravito-electro-magnetic vacuum with spacelike topology $R^2 \times
S^1$.
The amount of electromagnetic field that will be created ``from nothing"
is encoded in the initial state by a vector potential $A$, with $F=dA=0$,
but $\oint\! A$ non-vanishing around the $S^1$-direction. (This line integral
has no local effect classically, but could in principle be measured by an
Aharonov-Bohm interference between beams that traverse the $S^1$ in opposite
directions.) If we slice up the bounce solution that describes the decay by
a suitable set of 3-surfaces of constant Euclidean time,$^8$ we find at large
negative times a geometry that is still nearly flat, and an electromagnetic
field close to one of the run-away solutions of flat space: There is a small
but increasing $E$-field in the $S^1$-direction. Its displacement current
generates closed circles of $B$-field lines in the orthogonal plane ($R^2$).
This likewise increasing $B$-field induces further $E$-field, and so on. In
flat space this would eventually lead to infinite fields, say at $\tau =0$. But
gravity simultaneously exhibits the Witten$^6$ instability: a ``hole" in
space grows outward from a point at the center of the $B$-field circles. It
pushes these field lines outward, which induces an $E$-field opposing the one
originally present and reducing it to zero at the turn-around surface
$\tau =0$. On this surface the hole has reached its maximum size and is
enclosed by the maximum $B$-field. If developed further in Euclidean time, the
remaining history of the bounce reverses this time development, with the
$B$-field decreasing, the $E$-field temporarily reappearing, and the hole
vanishing, back to the vacuum state. But if we switch to Lorentzian time at the
turn-around surface, the state on that surface is the initial state
of the decay product, in which the hole later expands to infinity.

Another way to avoid the restrictions of the positive energy theorem is to
start with a state other than the vacuum, so that background fields are present
initially. Again pure Einstein gravity provides no truly static examples, but
because Maxwell fields can be repulsive, static Maxwell-Einstein configurations
do exist. But the repulsion acts only in two directions perpendicular to the
field direction, so the examples we consider are homogeneous and extend to
infinity in the field direction. (They would not be globally static, but
collapse in the field direction, if made finite in that direction by
identification via a discrete subgroup of the homogeneity.)

One such example is the Melvin universe, the closest analogy to a constant
magnetic field allowed in General Relativity. Garfinkle and Strominger$^9$ have
found a bounce solution that connects the Melvin universe with a state that is
asymptotically Melvin, but contains mouths of a maximally magnetically charged
wormhole, which are then pulled apart by the global magnetic field --- a pair
creation of magnetic wormhole charges. When we slice this bounce by 3-surfaces
to reveal the history of the tunneling, we can imagine the progress of the
field topology as follows: Assume the original $B$-field is vertical. Cut a
thin horizontal lens-shaped region out of space, but keep the top and bottom
surfaces of the lens identified (so the topology has not changed). The lens
intercepts some magnetic flux, which is subsequently kept constant. Blow up the
lens to a sphere with top and bottom hemispheres identified, and linking the
same flux. Then pinch the sphere inward at the equator until it pinches off
into
two spheres, changing the topology at the moment of pinch-off, but keeping the
linked flux unchanged. Now you have two spheres that are identified and link
the
flux --- they are the wormhole mouths created at the turning surface. The rest
of the bounce retraces this history back to the original Melvin state. This
gives only a very crude picture of what is happening to the geometry and the
field. (The exact solution of course contains all the details, but it is given
in coordinates that require some ingenuity in finding a suitable 3+1 split.)

Another example of a static Maxwell-Einstein space is the Bertotti-Robinson
uni\-verse,$^{10}$ which is not only homogeneous in the $B$-field ($z$-)
direction,
but constant curvature spherical in the two orthogonal (transverse) spacelike
directions. The Einstein equations then require that the $z,t$ subspace be
of constant curvature. An instanton solution has been given$^{11}$ that
connects
this with several Bertotti-Robinson universes. In a 3+1 spit that appears
natural the $E$-field vanishes on all slices. Thus this solution illustrates
the interaction of Euclidean gravity and $B$-field, rather than Euclidean
Maxwell theory. Again the history begins as a run-away solution: Near
$\tau = - \infty$ suppose the $B$-field is a little weaker near the equator of
the transverse $S^2$ than at the poles. This allows gravity to pinch in the
sphere at the equator, forcing more of the flux to the poles. The equator
pinches off at $\tau = 0$, forming two distorted spheres, each with more flux
in one half than the other. The flux and curvature even out, and the two
spheres become round at $\tau = \infty$. (The details of this are to be worked
out to obtain a mini-superspace in which to solve the Wheeler-DeWitt equation
and compare its results with those from the instanton.)

The duality transformation discussed in Section 3 allows us to change the
$B$-field, in terms of which all these examples were discussed, into an
$E$-field. For example, an electric Melvin universe will pair create
electrically charged wormholes. The first and third examples have also been
generalized to low-energy string theory.$^{7,12}$
\bigskip
\noindent {\bf 5. Multitudes of Universes}
\medskip
In this final section we want to consider Euclidean Maxwell-Einstein solutions
with more than one turning surface. Whether these can be interpreted as
tunneling solutions is much less clear than those of the previous sections. If
such interpretation is valid, then apparently the initial state, corresponding
to one of the turning surfaces, has been (carefully) prepared to be
time-symmetric, and the tunneling occurs when it has reached this moment of
time symmetry. The examples concern universe models, which typically reach
time-symmetry at the moment of maximum expansion --- an unlikely era for
quantum effects to play an important role. Nonetheless we describe these
examples briefly, because they are extensions to 4-dimensional Maxwell-Einstein
theory of an interesting geometrical technique developed for 2+1 gravity with
a negative cosmological constant.$^{13}$

The Euclidean version of the Bertotti-Robinson solution is geometrically
$S^2 \times H^2$, where $S^2$ is the round 2-sphere with electromagnetic field
2-form proportional to its area 2-form, and $H^2$ is a space of constant
negative curvature. Rather than finding a Euclidean solution that is
asymptotic to this one (as in ref.~11), we take exactly this solution and find
turning surfaces (totally geodesic surfaces) in it. These have the form
$S^2 \times$ geodesic of $H^2$. Now take a geodesic equilateral triangle in
$H^2$ and move the corners to infinity. The sides are then infinitely long
geodesics traversing the finite part of $H^2$, and they enclose a finite area.
The 4-space enclosed by the corresponding turning surfaces, regarded as an
instanton, therefore has finite action.\footnote*{The action should also
include contributions from the 2-boundaries at infinity$^{14}$, but these are
also finite.} In the spirit of ref.~13 this instanton may describe the break-up
of one Bertotti-Robinson universe into two, not by a transverse splitting as in
Section 4, but by ``snapping" longitudinally, like a rubber band that has been
stretched too much. Or, for that matter it might be viewed as a creation of a
triplet of universes from ``nothing," the ``nothing" being the $S^2$ that
carries the flux. (This construction can be generalized to involve more
universes; it is amusing to observe that creation from
``nothing" in this way must produce at least three universes!)

As another example, consider a geodesic regular hexagon in $H^2$. Because of
the negative curvature it is possible to make it of such a size that all the
interior angles are 90$^\circ$. Lay an identical copy on top and sew
together between top and bottom along sides 1, 3 and 5. Because the
corresponding surfaces $S^2 \times$ geodesic are totally geodesic, this is a
smooth identification. The remaining sides 2, 4 and 6 then form circles $S^1$
that are smooth due to the 90$^\circ$ angles, and hence correspond to turning
surfaces of topology $S^2 \times S^1$. If we follow the reasoning of ref.~13,
one of these, say from side 2, would be the initial state of a tunneling that
leads to two copies of $S^2 \times S^1$ (sides 4 and 6).

It is clear that these examples can be extensively generalized to involve a
multitude of universes. This shows at least that there exist Euclidean
solutions connecting widely different topologies in a 4-dimensional theory that
is usually regarded as reasonable. Lou Witten indeed chose to work on a
fascinating theory --- it can still give us interesting lessons and even
surprises!
\bigskip
\noindent {\bf 6. Acknowledgements}
\medskip
This research was supported in part by the National Science Foundation. I
appreciate the hospitality of the Aspen Center for Physics in Summer 1992,
where this paper was written.
\bigskip
\noindent {\bf 7. References}
\medskip
\item{1.} See, for example, L. Witten in {\it Gravitation, an Introduction to
Current
Research}, ed. L. Witten (John Wiley, New York 1962).

\item{2.} D. McLaughlin, {\it J. Math. Phys.} {\bf 13} (1972) 1099.

\item{3.} J.~J.~Halliwell and J.~B.~Hartle, {\it Phys. Rev.} {\bf D41} (1990)
1815;
G. W. Gibbons and J. B. Hartle, {\it Phys.~Rev.} {\bf D42} (1990) 2458;
S A Hayward, {\it Class. Quantum Grav.} {\bf 9} (1992) 1851.

\item{4.} S. Coleman, {\it Phys.~Rev.} {\bf D15} (1977) 2929; C. Callan and S.
Coleman, {\it ibid.} {\bf D16} (1977) 1762. For a recent critical discussion of
this method and further references, see D. Boyanowski et al, {\it Nucl. Phys.}
{\bf B376} (1992) 599.

\item{5.} K. Lee, {\it Phys.~Rev.~Lett.} {\bf 61} (1988) 263; S. Coleman and
K. Lee, {\it Nucl. Phys.} {\bf B329} (1990) 387.

\item{6.} E. Witten, {\it Commun. Math. Phys.} {\bf 80} (1981) 381.

\item{7.} D. Brill and G. Horowitz, {\it Phys.~Lett.} {\bf B262} (1991) 437.

\item{8.} For an example of such a slicing of the solution of ref.~6, see
D. Brill, {\it Found. Phys.} {\bf 16} (1986) 637.

\item{9.} D. Garfinkle and A. Strominger, {\it Phys.~Lett.} {\bf B256}
(1991) 146.

\item{10.} T. Levi-Civita, {\it R.C. Accad. Lincei} {\bf 26} (1917) 519;
B. Bertotti,
{\it Phys.~Rev.} {\bf 116} (1959) 1331; I. Robinson, {\it Bull. Akad. Polon.}
{\bf 7} (1959) 351.

\item{11.} D. Brill, {\it Phys.~Rev.} {\bf D46} (1992) 1560.

\item{12.} R. Kallosh, private communication, Aspen 1992.

\item{13.} Y. Fujiwara et al, {\it Prog. Theor. Phys.} {\bf 87} (1992) 253;
{\it Phys.~Rev.} {\bf D44} (1991) 1756 and 1763.

\item{14.} G. Hayward, {\it 2-boundary correction to the four dimensional
gravitational action}, University of British Columbia report (1992).
\end